\documentclass[preprintnumbers,amsmath,amssymb,superscriptaddress]{revtex4-2}

\usepackage{mathrsfs}
\usepackage{amssymb}
\usepackage{graphicx}
\usepackage{dcolumn}
\usepackage{bm}
\usepackage{textcomp}
\usepackage{amsmath}
\usepackage[titletoc]{appendix}
\usepackage{accents}
\usepackage{ntheorem}
\usepackage{color}

\newcommand\ket[1]{\ensuremath{|#1\rangle}}

\newcommand\oprod[2]{\ensuremath{|#1\rangle\langle#2|}}
\newcommand\mean[1]{\ensuremath{\langle #1\rangle}}

\newcounter{RomanNumber}

\makeatletter
\def\widebar{\accentset{{\cc@style\underline{\mskip10mu}}}}
\def\Widebar{\accentset{{\cc@style\underline{\mskip8mu}}}}
\makeatother

\begin{document}

\title{Robust twin-field quantum key distribution through sending-or-not-sending}

\author{Cong Jiang}
\affiliation{Jinan Institute of Quantum Technology, Jinan, Shandong 250101, P.~R.~China}
\affiliation{State Key Laboratory of Low Dimensional Quantum Physics, Department of Physics, Tsinghua University, Beijing 100084, P.~R.~China}

\author{Zong-Wen Yu}

\affiliation{Data Communication Science and Technology Research Institute, Beijing 100191, P.~R.~China}

\author{Xiao-Long Hu}
\affiliation{School of Physics, State Key Laboratory of Optoelectronic Materials and Technologies, Sun Yat-sen University, Guangzhou 510275, China}

\author{Xiang-Bin Wang}\email{Corresponding author: xbwang@mail.tsinghua.edu.cn}
\affiliation{Jinan Institute of Quantum Technology, Jinan, Shandong 250101, P.~R.~China}
\affiliation{State Key Laboratory of Low Dimensional Quantum Physics, Department of Physics, Tsinghua University, Beijing 100084, P.~R.~China}
\affiliation{Shanghai Branch, CAS Center for Excellence and Synergetic Innovation Center in Quantum Information and Quantum Physics, University of Science and Technology of China, Shanghai 201315, P.~R.~China}
\affiliation{ Shenzhen Institute for Quantum Science and Engineering, and Physics Department, Southern University of Science and Technology, Shenzhen 518055, P.~R.~China}
\affiliation{Frontier Science Center for Quantum Information,
Beijing, P.~R.~China}

\begin{abstract}
The sending-or-not-sending (SNS) protocol is one of the most major variants of the twin-field (TF) quantum key distribution (QKD) protocol and has been realized in a 511 km field fiber, the farthest field experiment to date. In practice, however, all decoy-state methods have unavoidable source errors, and the source errors may be non-random, which compromises the security condition of the existing TF-QKD protocols. In this study, we present a general approach for efficiently calculating the SNS protocol’s secure key rate with source errors, by establishing the equivalent protocols through virtual attenuation and tagged model. This makes the first result for TF-QKD in practice where source intensity cannot be controlled exactly. Our method can be combined with the two-way classical communication method such as active odd-parity pairing to further improve the key rate. The numerical results show that if the intensity error is within a few percent, the key rate and secure distance only decrease marginally. The key rate of the recent SNS experiment in the 511 km field fiber is still positive using our method presented here, even if there is $\pm 9.5\%$ intensity fluctuation. This shows that the SNS protocol is robust against source errors.
\end{abstract}


\maketitle
\section{Introduction}
Since the proposal of the twin-field (TF) quantum key distribution (QKD)~\cite{lu2018overcoming}, the longest distance record of QKD has been constantly and rapidly refreshed in recent years~\cite{liu2019experimental,wang2019beating,zhong2019proof,fang2020implementation,chen2020sending,liu2021field,chen2021twin,pittaluga2021600}. The upper bounds of the key rate of the previous protocols, such as the BB84 protocol~\cite{bennett1984quantum,hwang2003quantum,wang2005beating,lo2005decoy,lim2014concise,boaron2018secure} and the measurement-device-independent(MDI)-QKD protocol~\cite{lo2012measurement,braunstein2012side,wang2013three,curty2014finite,xu2014protocol,yu2015statistical,zhou2016making} are limited to the linear scale of channel transmittance, also known as the PLOB bound~\cite{pirandola2017fundamental}. Based on the single-photon interference, the TF-QKD can raise the key rate from linear scale to square root scale of channel transmittance and break the PLOB bound.

As one of the most important variants of TF-QKD~\cite{wang2018twin,tamaki2018information,cui2019twin,curty2018simple}, the sending-or-not-sending (SNS) protocol~\cite{wang2018twin} has been extensively studied in theories \cite{yu2019sending,jiang2019unconditional,hu2019general,xu2020sending,jiang2021composable,jiang2020zigzag} and experiments~\cite{liu2019experimental,chen2020sending,liu2021field,chen2021twin,pittaluga2021600,clivati2022coherent}. Specifically, the 511-km field experiment~\cite{chen2021twin} was done by applying the SNS protocol with actively odd-parity pairing (AOPP) method, and the 605-km laboratory experiment~\cite{pittaluga2021600} was done by applying the SNS protocol with standard two-way classical communication (TWCC) method. The 511-km experiment is the farthest field experiment to date. 

Given its great progress~\cite{pirandola2020advances,xu2020secure} in both theory and technology, more and more studies of QKD are pointing to real world application with various technology limitations.

The TF-QKD has typically used with the weak coherent state (WCS) sources. The decoy-state method must be used to ensure the security of TF-QKD with WCS sources. The intensities of the WCS sources are always assumed to be stable in the entire protocol, which is not the fact in experiments where source errors are unavoidable. The source errors, in particular, are not always random. For example, the intensities of the sources can slowly shift over time and are affected by temperature changes in the environment. 

Although there are results for decoy-state analysis with source errors for the BB84 and the MDI-QKD protocol~\cite{wang2008general,jiang2016measurement}, they do not solve the problem for the SNS protocol where encoding is done by the vacuum and single-photon states themselves. {To calculate the key rate, we need a way to estimate the single-photon phase-flip error rate.} It is not a trivial task because the single photon state in $X$ basis is now different from that in $Z$ basis. We shall solve this problem by virtual attenuation and then apply the existing theory for decoy-state analysis with intensity fluctuation.

The work is organized as follows: first, we introduce the real and virtual protocols and demonstrate their equivalence. This shows how to efficiently calculate the secure key rate in a practical application where intensities cannot be controlled exactly. Then we present some numerical results that compare the key rates of the SNS protocol with different degrees of source errors.

\section{The security of SNS protocol with source errors}
For ease of presentation, here we introduce the SNS protocol through the 4-intensity model. The scientific content is the same as in the earlier literatures.

We take the 4-intensity SNS protocol as an example to show the security proof. There are four sources with different intensities on Alice's and Bob's sides. We denote Alice's sources by $a_0,a_1,a_2$ and $a_z$, and Bob's sources by $b_0,b_1,b_2$ and $b_z$, where $a_0$ and $b_0$ are the vacuum sources. WCS sources are typically used to perform the SNS protocol. If the sources are stable, {the intensities of the sources $a_l$ and $b_r$ are $\mu_{a_l}$ and $\mu_{b_r}$, respectively, for $l,r=0,1,2,z$} where $\mu_{a_0}=\mu_{b_0}=0$. However, in practice, the sources are always unstable and the intensities are different in different time windows. The intensities of the sources in the $i$-th time window are denoted as $\mu_{a_l}^i$ and $\mu_{b_r}^i$ respectively, where
\begin{equation}
\mu_{a_l}^i=(1+\delta_{a_l}^i)\mu_{a_l}, \quad \mu_{b_r}^i=(1+\delta_{b_r}^i)\mu_{b_r}
\end{equation}
and $|\delta_{a_l}^i|\le \delta_{a_l},|\delta_{b_r}^i|\le \delta_{b_r}$. We assume $\delta_{a_l}$ and $\delta_{b_r}$ are known values in the protocol. The lower and upper bounds of a physical quantity are represented by superscripts $L$ and $U$, respectively. For instance, $\mu_{a_1}^U$ is the upper bound of the intensity of the source $a_1$, and we have $\mu_{a_1}^U=(1+\delta_{a_1})\mu_{a_1}$.

\subsection{The real protocol}\label{realprotocol}
Alice (Bob) randomly decides whether the $i$-th time window is a decoy window or a signal window with probabilities $1-p_{a_z}$ and $p_{a_z}$ ($1-p_{b_z}$ and $p_{b_z}$). If it is a decoy window, Alice (Bob) randomly chooses the sources $a_0,a_1,a_2$ ($b_0,b_1,b_2$) with probabilities $p_{a_0}=1-p_{a_1}-p_{a_2},p_{a_1},p_{a_2}$ ($p_{b_0}=1-p_{b_1}-p_{b_2},p_{b_1},p_{b_2}$). Since the sources are unstable, Alice (Bob) actually prepares a WCS pulse in state $\ket{e^{\imath\theta_{a_l}^i}\sqrt{\mu_{a_l}^i}}$ ($\ket{e^{\imath\theta_{b_r}^i}\sqrt{\mu_{b_r}^i}}$) if the source $a_l$ ($b_r$) is chosen for $l,r=1,2$, where $\theta_{a_l}$ and $\theta_{b_r}$ are random in $[0,2\pi)$. The imaginary unit is represented here by the symbol $\imath$. If it is a signal window, Alice (Bob) randomly chooses the sources $a_0,a_z$ ($b_0,b_z$) with probabilities $1-\epsilon_a,\epsilon_a$ ( $1-\epsilon_b,\epsilon_b$). Alice (Bob) actually prepares a phase-randomized WCS pulse with intensity $\mu_{a_z}^i$ ($\mu_{b_z}^i$) if the source $a_z$ ($b_z$) is chosen. {Clearly, if Alice (Bob) commits to a signal window at time $i$, Alice's (Bob's) choice of source $a_z$ ($b_z$) indicates a decision of \emph{sending} whereas the choice of source $a_0$ ($b_0$) is a decision of \emph{not-sending} in SNS protocol.}

{\emph{They} (Alice and Bob)} send the prepared pulse pair to Charlie, who is assumed to perform interferometric measurements on the received pulse pair. Then Charlie announces the measurement results to \emph{them}. If only one detector clicks, \emph{they} would take it as a \emph{one-detector heralded event}. After \emph{they} repeat the above process for $N$ times and Charlie announces all measurement results, \emph{they} acquire a series of data. Then \emph{they} perform the data post processing, including post selection of events in $X$ windows and final key distillation.

\emph{They} first announce the type of each time window they decide, i.e., whether it is a decoy window or a signal window. For a time window that both of \emph{them} decide a signal window, it is a $Z$ window. The one-detector heralded events in $Z$ windows are effective events, and the corresponding bits of those effective events formed the $n_t$-bit raw key strings, which are used to extract the final keys. We also define the $\tilde{Z}$ window as a $Z$ window when sources $\{a_0, b_z\}$ or $\{a_z, b_0\}$ are used, i.e., {a $Z$ window when one side decides sending and the other side decides not-sending}. The untagged pulses in the $Z$ windows are that Alice decides not sending and Bob actually sends out a single-photon pulse or Bob decides not sending and Alice actually sends out a single-photon pulse, i.e., the single-photon pulses in the $\tilde{Z}$ windows. The untagged bits are the corresponding bits of those untagged pulses {that cause effective events}. Except the $Z$ windows, \emph{they} announce the intensities of the pulses in each window, and for the time windows that both of \emph{them} decide sending out a pulse of source $a_1$ and $b_1$ respectively, \emph{they} also announce the phases of the pulse pairs. For a time window that both of \emph{them} decide sending out a pulse of source $a_1$ and $b_1$ respectively, and their phases satisfy the post selection criteria~\cite{wang2018twin}, it is defined as an $X$ window. The one-detector heralded events in $X$ windows are effective events. The effective events in $X$ windows are used to estimate the phase-flip error rate.

According to Ref.~\cite{hu2019general}, if the sources are stable, and source $a_z$ ($b_z$) always emits a state of $\rho_{a_z}=\sum_{k=0} q_A^k \oprod{k}{k}$,
 ($\rho_{b_z}=\sum_{k=0} q_B^k \oprod{k}{k}$), we can use the following constraint for the source for security, 
\begin{equation}\label{secur_con}
\frac{\mu_{a_1}}{\mu_{b_1}}=\frac{\epsilon_a(1-\epsilon_b)q_A^1}{\epsilon_b(1-\epsilon_a)q_B^1},
\end{equation}
Here $\oprod{k}{k}$ represents for a $k$-photon Fock state. In particular, when using a phase randomized WCS source with intensity $\mu_{a_z}$ ($\mu_{b_z}$) for source $a_z$ ($b_z$), $q_A^1,q_B^1$ in Eq.~\eqref{secur_con} can be replaced by $\mu_{a_z}e^{-\mu_{a_z}},\mu_{b_z}e^{-\mu_{b_z}}$. Note that if the constraint above is respected, the density matrix of single-photon state in $\tilde{Z}$ windows is identical to that of $X$ basis and hence we can faithfully verify the phase-flip error rate of untagged bits by observing events in $X$ windows.

Given stable sources with the source parameters known, the SNS protocol is still secure and efficient even if Eq.~\eqref{secur_con} does not hold. Most straightforwardly, we can post-select the sending probabilities $\epsilon_a,\epsilon_b$ by randomly deleting some of the time windows~\cite{chen2020sending}, i.e., to make Eq.~\eqref{secur_con} hold by post-deletion. (Post deleting a time window means disregarding the event happening at that time window.) Although the amount of data is reduced by the post-deletion, the key rate will essentially be unchanged if the original bias between the two sides of Eq.~\eqref{secur_con} is not too large, because, after post deletion, the error rate of raw keys is also decreased. 

The more practical situation is that we do not know the exact intensities of the sources of the Z windows but we know the bound values of intensities for each source of Z windows. In such a case, we can use the tagged model: replace the intensities of Eq.~\eqref{secur_con} by a lower bound value and make Eq.~\eqref{secur_con} hold. 

Most generally, the source intensities of $X$ windows are also unstable. Say, in practice, none of the sources are stable, and the intensity errors may be different from time to time and not necessarily random. In such a case, Eve can treat different time windows differently, thus we can not treat the $X$ windows as a whole, but have to deal with each time window separately. The intensities of sources {$a_1,b_1$} in different time windows are different, which means in general
\begin{equation}
\frac{\mu_{a_1}^i}{\mu_{b_1}^i}\neq \frac{\mu_{a_1}^j}{\mu_{b_1}^j},
\end{equation}
if $i\neq j$. When Eq.~\eqref{secur_con} holds, the bit-flip error rate of $X$ windows can be used to estimate the phase-flip error rate of the untagged bits in $Z$ windows. But due to the source errors, Eq.~\eqref{secur_con} can not be held for all $X$ windows, so that we can not estimate the phase-flip error rate of the untagged bits according to the data in $X$ windows. {The situation is different for the decoy-state BB84 protocol with source errors~\cite{wang2008general}, or the decoy-state MDI-QKD~\cite{lo2012measurement,zhou2016making} where the source errors do not change the encoding states in $X$ windows or $Z$ windows, it affects the decoy-state analysis only~\cite{wang2008general,jiang2016measurement}.} In this work, we solve this problem based on SNS protocol and makes the TF-QKD robust with imperfect real set-ups.

\subsection{Main idea: Quantum security from virtual attenuation}
The state preparation stage of the real protocol can be regarded as this: At every time window $i$, Alice (Bob) has 4 candidate states emitted from the 4 sources $a_0,a_1,a_2,a_z$ ($b_0,b_1,b_2,b_z$), respectively. In particular, the intensities of the coherent states emitted by source $a_l, b_r$ are $\mu_{a_l}^i,\mu_{b_r}^i$ for $l,r=1,2,z$, respectively. Sources $a_0,b_0$ emit vacuum only. Alice (Bob) will choose one state from the $4$ candidate states for the time window $i$, by the probabilities stated earlier in the real protocol.

Our first major idea is mapping by (virtual) attenuation~\cite{wang2007simple}. We regard the candidate states above as the attenuation outcome from virtual sources $a_0,a_1^\prime,a_2^\prime,a_z^\prime$ ($b_0,b_1^\prime,b_2^\prime,b_z^\prime$) and the intensity of {the WCS pulse} from virtual source $a_1^\prime$ ($b_1^\prime$) is fixed to $\mu_{a_1}^U$ ($\mu_{b_1}^U$) exactly at any time window $i$. This naturally means that the transmittance $\eta_A^i=\frac{1+\delta_{a_1}^i}{1+\delta_{a_1}}$ ($\eta_B^i=\frac{1+\delta_{b_1}^i}{1+\delta_{b_1}}$) in the virtual attenuation of the $i$-th time window, and hence the intensities of {the WCS pulses} from virtual sources $a_2^\prime,a_z^\prime$ ($b_2^\prime,b_z^\prime$) are $\mu_{a_2}^{\prime i}= \mu_{a_2}^{i}/\eta_A^i, \mu_{a_z}^{\prime i}= \mu_{a_z}^{i}/\eta_A^i$, ($\mu_{b_2}^{\prime i}= \mu_{b_2}^{i}/\eta_B^i,\mu_{b_z}^{\prime i}= \mu_{b_z}^{i}/\eta_B^i$), respectively. For simplicity, we shall just call virtual protocol (real protocol) for the SNS protocol using virtual (real) sources above. Surely, the real protocol above is secure provided that the virtual protocol above is secure. This holds even though Eve knew the exact values of intensity fluctuation of each source in every time window $i$. Or equivalently, even though Eve knew the exact values of $\eta_a^i$, $\eta_b^i$. If Eve can use scheme $\mathcal{G}$ to attack the real protocol, {Eve} can also attack the virtual protocol by first taking attenuation $\{\eta_A^i,\eta_B^i\}$ to pulses in the virtual protocol {and then use attacking scheme $\mathcal{G}$}. Surely, we can obtain the secure key rate for the real protocol above by calculating the secure key rate of the virtual protocol, because the two protocols cause no difference to outside lab. 

Our second major idea is to treat the intensity fluctuation for states of the $Z$ windows by the tagged model. Consider the key rate calculation for our virtual protocol above. In that protocol, at time window $i$, the intensity of the pulse from virtual source {$a_z^\prime$ ($b_z^\prime$)} is $\mu_{a_z}^\prime$ ($\mu_{b_z}^\prime$). This in general does not respect Eq.~\eqref{secur_con}. However, in $\tilde{Z}$ windows, there does exist a subset of single-photon pulses which are indistinguishable from the single-photons in $X$ windows, and there does exist an efficient method to verify the lower bound of the size of this subset. Thus the key rate can be effectively calculated.

The third major idea here is to verify the lower bound values of the untagged bits in $Z$ windows and their corresponding upper bound value of phase flip error rate, by applying the existing decoy-state method with source errors~\cite{wang2008general}. Note that in the virtual protocol above, the intensities of the stronger decoy pulses are not stable. According to the existing theory, we only need to use its lower bound values of the photon number distribution coefficients to obtain the worst-case result. With these, the final key rate can be calculated. Furthermore, we can apply the TWCC method such as AOPP to improve the key rate~\cite{xu2020sending,jiang2021composable}.

\subsection{The virtual protocol 1}\label{vrp1}
In the $i-$th time window of virtual protocol 1, the pulses emitted by sources $a_1,a_2,a_z$ ($b_1,b_2,b_z$) are different from those of the real protocol. To avoid confusion, we use notations $a_0,a_1^\prime,a_2^\prime,a_z^\prime$ ($b_0,b_1^\prime,b_2^\prime,b_z^\prime$) for Alice’s (Bob's) sources in virtual protocol 1. 

In the $i$-th time window, Alice (Bob) randomly decides it is a decoy window or a signal window with probabilities $1-p_{a_z}$ and $p_{a_z}$ ($1-p_{b_z}$ and $p_{b_z}$). If it is a decoy window, Alice (Bob) randomly chooses the sources $a_0,a_1^\prime,a_2^\prime$ ($b_0,b_1^\prime,b_2^\prime$) with probabilities $p_{a_0}=1-p_{a_1}-p_{a_2},p_{a_1},p_{a_2}$ ($p_{b_0}=1-p_{b_1}-p_{b_2},p_{b_1},p_{b_2}$). If the source $a_1^\prime$ ($b_1^\prime$) is chosen, Alice (Bob) actually prepares a WCS pulse in state $\ket{e^{\imath\theta_{a_1}^i}\sqrt{\mu_{a_1}^U}}$ ($\ket{e^{\imath\theta_{b_1}^i}\sqrt{\mu_{b_1}^U}}$), where $\mu_{a_1}^U=(1+\delta_{a_1})\mu_{a_1}$ and $\mu_{b_1}^U=(1+\delta_{b_1})\mu_{b_1}$. If the source $a_2^\prime$ ($b_2^\prime$) is chosen, Alice (Bob) actually prepares a WCS pulse in state $\ket{e^{\imath\theta_{a_2}^i}\sqrt{\mu_{a_2}^{\prime i}}}$ ($\ket{e^{\imath\theta_{b_2}^{i}}\sqrt{\mu_{b_2}^{\prime i}}}$), where
\begin{equation}
\mu_{a_2}^{\prime i}= \frac{1+\delta_{a_1}}{1+\delta_{a_1}^i} \mu_{a_2}^{i}, \quad \mu_{b_2}^{\prime i}= \frac{1+\delta_{b_1}}{1+\delta_{b_1}^i} \mu_{b_2}^{i}.
\end{equation}
If it is a signal window, Alice (Bob) randomly chooses the sources $a_0,a_z^\prime$ ($b_0,b_z^\prime$) with probabilities $1-\epsilon_a,\epsilon_a$ ( $1-\epsilon_b,\epsilon_b$). Alice (Bob) actually prepares a phase-randomized WCS pulse with intensity $\mu_{a_z}^{\prime i}$ ($\mu_{b_z}^{\prime i}$) if the source $a_z^\prime$ ($b_z^\prime$) is chosen, where
\begin{equation}
\mu_{a_z}^{\prime i}= \frac{1+\delta_{a_1}}{1+\delta_{a_1}^i} \mu_{a_z}^{i}, \quad \mu_{b_z}^{\prime i}= \frac{1+\delta_{b_1}}{1+\delta_{b_1}^i} \mu_{b_z}^{i}.
\end{equation}

Then \emph{they} send out the prepared pulse pair. There are two attenuators between the channel and Alice's and Bob's labs, which are controlled by David. To clarify, we define the attenuator on Alice's side as attenuator A and the attenuator on Bob's side as attenuator B. In the $i-$th time window, David sets the transmittance of the attenuator A to $\eta_A^i=\frac{1+\delta_{a_1}^i}{1+\delta_{a_1}}$, and the transmittance of the attenuator B to $\eta_B^i=\frac{1+\delta_{b_1}^i}{1+\delta_{b_1}}$. Then we have
\begin{equation}\label{equiv1}
\begin{split}
&\mu_{a_1}^i=\eta_A^i\mu_{a_1}^U,\quad \mu_{a_2}^{i}=\eta_A^i \mu_{a_2}^{\prime i}, \quad \mu_{a_z}^{i}=\eta_A^i \mu_{a_z}^{\prime i},\\
&\mu_{b_1}^i=\eta_B^i\mu_{b_1}^U,\quad \mu_{b_2}^{i}=\eta_B^i \mu_{b_2}^{\prime i}, \quad \mu_{b_z}^{i}=\eta_B^i \mu_{b_z}^{\prime i}.
\end{split}
\end{equation}

\begin{figure}
\centering
\includegraphics[width=9 cm]{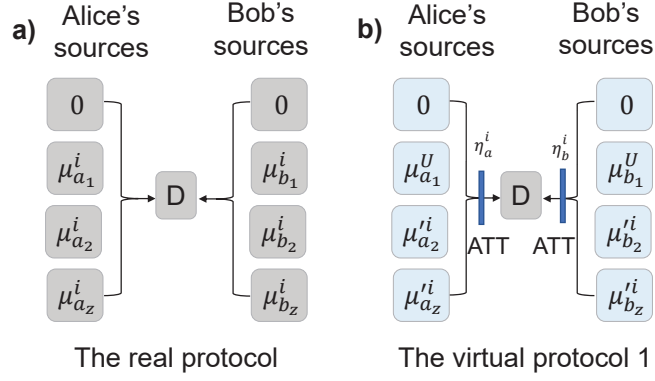}
\caption{The comparison of the real protocol and the virtual protocol 1. Here the box `D' represents Charlie‘s detectors and the `ATT' represents the attenuator. Although the sources of the real protocol and the virtual protocol 1 are different, Eve cannot distinguish the differences due to the proper set attenuators.}\label{virtual_protocol}
\end{figure}

Eq.~\eqref{equiv1} means the real protocol and virtual protocol 1 are equivalent to Eve who is assumed to control the channel and detectors. Thus the information leakages in the real protocol and virtual protocol 1 are the same and we can use virtual protocol 1 to estimate the key rate of the real protocol. A simple comparison of the real protocol and virtual protocol 1 is shown in Figure.~\ref{virtual_protocol}.

Note in virtual protocol 1, the sources $a_1^\prime$ and $b_1^\prime$ are stable, bringing us one step closer to the final goal. However, due to the unstable sources in the $Z$ windows, the security condition Eq.~\eqref{secur_con} is not always satisfied. Recall that the untagged pulses in the $Z$ windows are the single-photon pulses in the $\tilde{Z}$ windows. This means that we should only need to care about the single-photon pulse of the sources $a_z^\prime$ and $b_z^\prime$. The density matrix of a phase-randomized WCS pulse with intensity $\mu$ is
\begin{equation}
\rho_{\mu}=\sum_{k=0}^{+\infty}\frac{\mu^ke^{-\mu}}{k!}\oprod{k}{k}.
\end{equation}    
  
Although the source $a_z^\prime$ is unstable, the density matrix of the pulse from this source in the $i$-th time window can be expressed in the following convex form
\begin{equation}\label{azi1}
\rho_{\mu_{a_z}^{\prime i}}=c_{a_z}^1\oprod{1}{1}+(1-c_{a_z}^1)\rho_{a_c}^i,
\end{equation}
where 
\begin{equation}
\rho_{a_c}^i=\frac{1}{1-c_{a_z}^1}[(\mu_{a_z}^{\prime i}e^{-\mu_{a_z}^{\prime i}}-c_{a_z}^1)\oprod{1}{1}+e^{-\mu_{a_z}^{\prime i}}\oprod{0}{0}+\sum_{k=2}^{+\infty}\frac{(\mu_{a_z}^{\prime i})^ke^{-\mu_{a_z}^{\prime i}}}{k!}\oprod{k}{k}],
\end{equation}
for $c_{a_z}^1\le \mu_{a_z}^{\prime i}e^{-\mu_{a_z}^{\prime i}}$. And if $c_{a_z}^1\le \mu_{a_z}^{\prime L}e^{-\mu_{a_z}^{\prime L}}$, the density matrices of all pulses of source $a_z^\prime$ have the similar convex form as shown in Eq.~\eqref{azi1}, where the only differences are $\rho_{a_c}^i$ for different time windows. But as stated before, we only need to care about the single-photon pulse of the source $a_z^\prime$, thus {as long as} $c_{a_z}^1\le \mu_{a_z}^{\prime L}e^{-\mu_{a_z}^{\prime L}}$, we can take the unstable source $a_z^\prime$ as the mixture of two sources: one stable source $a_z^{\prime 1}$ with probability $c_{a_z}^1$ and one unstable source $a_z^{\prime c}$ with probability $1-c_{a_z}^1$, where the source $a_z^{\prime 1}$ only emits perfect single-photon pulses and the source $a_z^{\prime c}$ emits {pulses in state} $\rho_{a_c}^i$. 

For the unstable source $b_z^\prime$, the density matrix of the pulse in $i-$th time window is
\begin{equation}\label{bzi1}
\rho_{\mu_{b_z}^{\prime i}}=c_{b_z}^1\oprod{1}{1}+(1-c_{b_z}^1)\rho_{b_c}^i,
\end{equation}
where 
\begin{equation}
\begin{split}
&\rho_{b_c}^i=\frac{1}{1-c_{b_z}^1}[(\mu_{b_z}^{\prime i}e^{-\mu_{b_z}^{\prime i}}-c_{b_z}^1)\oprod{1}{1}+\\
&e^{-\mu_{b_z}^{\prime i}}\oprod{0}{0}+\sum_{k=2}^{+\infty}\frac{(\mu_{b_z}^{\prime i})^ke^{-\mu_{b_z}^{\prime i}}}{k!}\oprod{k}{k}],
\end{split}
\end{equation}
for $c_{b_z}^1\le \mu_{b_z}^{\prime i}e^{-\mu_{b_z}^{\prime i}}$. As long as $c_{b_z}^1\le \mu_{b_z}^{\prime L}e^{-\mu_{b_z}^{\prime L}}$, the unstable source $b_z^\prime$ can be regarded as the mixture of two sources: one stable source $b_z^{\prime 1}$ with probability $c_{b_z}^1$ and one unstable source $b_z^{\prime c}$ with probability $1-c_{b_z}^1$ where source $b_z^{\prime 1}$ only emits perfect single-photon pulses and $b_z^{\prime c}$ emits {pulses in state} $\rho_{b_c}^i$. With this, we have the following virtual protocol 2.

\subsection{The virtual protocol 2}\label{virp2}
In the $i$-th time window, Alice (Bob) randomly decides it is a decoy window or a signal window with probabilities $1-p_{a_z}$ and $p_{a_z}$ ($1-p_{b_z}$ and $p_{b_z}$). If it is a decoy window, Alice (Bob) randomly chooses the sources $a_0,a_1^\prime,a_2^\prime$ ($b_0,b_1^\prime,b_2^\prime$) with probabilities $p_{a_0}=1-p_{a_1}-p_{a_2},p_{a_1},p_{a_2}$ ($p_{b_0}=1-p_{b_1}-p_{b_2},p_{b_1},p_{b_2}$). If it is a signal window, Alice (Bob) randomly chooses the sources $a_0,a_z^{\prime 1}, a_z^{\prime c}$ ($b_0,b_z^{\prime 1}, b_z^{\prime c}$) with probabilities $1-\epsilon_a,\epsilon_a c_{a_z}^1, \epsilon_a(1-c_{a_z}^1)$ ( $1-\epsilon_b,\epsilon_b c_{b_z}^1, \epsilon_b(1-c_{b_z}^1)$), where the source $a_z^{\prime 1}$ only emits single-photon pulses and the source $a_z^{\prime c}$ emits pulses in state $\rho_{a_c}^i$ (the source $b_z^{\prime 1}$ only emits single-photon pulses and the source $b_z^{\prime c}$ emits pulses in state $\rho_{b_c}^i$), and if 
\begin{equation}
\frac{\mu_{a_z}^{\prime L}e^{-\mu_{a_z}^{\prime L}}}{\mu_{b_z}^{\prime L}e^{-\mu_{b_z}^{\prime L}}}\le \frac{\epsilon_b(1-\epsilon_a)\mu_{a_1}^U}{\epsilon_a(1-\epsilon_b)\mu_{b_1}^U},
\end{equation}
 $c_{a_z}^1, c_{b_z}^1$ satisfy
\begin{equation}
\begin{split}
c_{a_z}^1=\mu_{a_z}^{\prime L}e^{-\mu_{a_z}^{\prime L}}, \quad c_{b_z}^1=\frac{\epsilon_a(1-\epsilon_b)\mu_{b_1}^U}{\epsilon_b(1-\epsilon_a)\mu_{a_1}^U}c_{a_z}^1,
\end{split}
\end{equation}
else
\begin{equation}
c_{b_z}^1=\mu_{b_z}^{\prime L}e^{-\mu_{b_z}^{\prime L}}, \quad c_{a_z}^1=\frac{\epsilon_b(1-\epsilon_a)\mu_{a_1}^U}{\epsilon_a(1-\epsilon_b)\mu_{b_1}^U}c_{b_z}^1.
\end{equation}

The following precesses are the same with the virtual protocol 1. As discussed above, the virtual protocol 2 is equivalent to the virtual protocol 1. Eve can not distinguish the difference between the real protocol, virtual protocol 1 and virtual protocol 2.

We shall regard all those (single-photon) pulses from sources $\{a_z^{\prime 1}, b_0\}$ and $\{a_0, b_z^{\prime 1}\}$ in $Z$ windows as untagged pulses. Clearly, a state of untagged pulse here is identical to the single-photon state in $X$ windows {due to}
\begin{equation}
\frac{\mu_{a_1}^U}{\mu_{b_1}^U}=\frac{\epsilon_a(1-\epsilon_b)c_{a_z}^1}{\epsilon_b(1-\epsilon_a)c_{b_z}^1}.
\end{equation}

Furthermore, only the sources $a_2^\prime$ and $b_2^\prime$ are unstable in virtual protocol 2. The remaining problems are estimating the lower bound of the number of untagged bits and the upper bound of the phase-flip error rate, both of which can be solved by directly applying the conclusion of Ref.~\cite{wang2008general}. Finally, using the method introduced in Ref.~\cite{yu2019sending,jiang2019unconditional}, we can obtain the final key rate. We can also use the AOPP method to improve the key rate~\cite{xu2020sending,jiang2021composable}. The calculation details are shown in Appendix~\ref{cal}.

\subsection{Major results}
1. Given the unstable sources as presented in the real protocol, we map them to virtual sources by virtual attenuation {and tagged model}, as shown in the virtual protocol 2. 
 
2. In the virtual protocol 2, the source $a_1^\prime$ and $b_1^\prime$ are stable,
 with exactly known intensities, and the sources $a_z^{\prime 1}$ and $b_z^{\prime 1}$ are perfect single-photon sources, while other non-vacuum sources are unstable, with known bounds.
 
3. Regard the observed values in the real protocol to be those of the virtual protocol, and use decoy-state analysis to verify the value $n_1^L$ and $e_1^{ph,U}$, which are the lower bound value of the number of untagged bits in ${Z}$ windows and the upper bound value of the phase-flip error rate.

4. Calculate the key rate for the real protocol by the following formula:
\begin{equation}\label{rr2}
\begin{split}
R=\frac{1}{N}\{n_1^L[1-H(e_1^{ph,U})]-fn_tH(E)-\log_2 \frac{2}{\varepsilon_{cor}}-2\log_2\frac{1}{\sqrt{2}\varepsilon_{PA}\hat{\varepsilon}}\}.
\end{split}
\end{equation}
where $N$ represents the total number of pulse pairs sent by Alice and Bob, $f$ represents the error correction inefficiency, $n_t$ represents the number of effective events in the $Z$ windows, and $E$ represents the error rate of the raw keys in the $Z$ windows. As shown in Ref.~\cite{curty2014finite,jiang2019unconditional}, the tailing term of $-\log_2 \frac{2}{\varepsilon_{cor}}-2\log_2\frac{1}{\sqrt{2}\varepsilon_{PA}\hat{\varepsilon}}$ is the additional cost for security with finite size. The calculation details are shown in Appendix~\ref{cal}. {As detailed in Appendix~\ref{decoy}, our method does not presume the source errors to be random. Our protocol assumes that Eve can know or even determine the error values of all those candidate states in advance, so they surely are not limited to be random errors only. For security, the decoy state method requests that Eve has no information in advance about secret state choice of Alice and Bob, i.e., choosing the pulse of which source to send out by Alice and Bob for each time windows. To keep this condition, our method requests that the values of intensity errors which can be known to Eve in advance do not carry any private information of state choice of Alice and Bob. With this presumption being respected, our method allows whatever dependence of intensity errors from different time windows.}

5. The AOPP method is directly applicable here. Suppose Bob generates the active odd-parity random pairing and then informs Alice of the positions of two bits in raw keys for each pair. Following the parity check, all pairs with odd-parity values at Alice's side survive, and one bit from each survived pair is chosen at random for final key distillation. That is, in the AOPP, a pair with two untagged bits will always contribute an untagged bit to the final key distillation. Those pairs with even parity values on Alice's side will be completely discarded. We have the following final key length formula after AOPP
\begin{equation}\label{r2}
\begin{split}
R^\prime=\frac{1}{N}\{n_1^{\prime L}[1-H(e_{1}^{\prime ph,U})]-fn_t^\prime H(E^\prime)-2\log_2{\frac{2}{\varepsilon_{cor}}}-4\log_2{\frac{1}{\sqrt{2}\varepsilon_{PA}\hat{\varepsilon}}}\}.
\end{split}
\end{equation}
where $n_1^{\prime L}$ is the lower bound of untagged bits after AOPP; $e_{1}^{\prime ph,U}$ is the upper bound of the phase-flip error rate after AOPP; $n_t^\prime$ is the number of survived bits after AOPP; $E^\prime$ is the bit-flip error rate of the survived bits after AOPP. Details are shown in Appendix~\ref{cal}.

\section{Numerical simulation}
We use the linear model to simulate the observed values~\cite{yu2019sending}, and Charlie’s detectors are assumed to be identical, that is, they have the same detection efficiency and dark counting rate. For simplicity, we assume the maximum derivations of all sources are the same, i.e., $\delta_{a_l} = \delta_{b_r} = \delta$ for $l,r = 1,2,z$. The experiment parameters are listed in Table.~\ref{exproperty}.

\begin{table*}[htbp]
\centering
\begin{tabular}{ccccccc}
\hline
$p_d$& $e_d$ &$\eta_d$ & $f$ & $\alpha_f$ &$\xi$ &$N$\\
\hline
$1.0\times10^{-8}$& {$4\%$}  & {$60.0\%$} & $1.1$ & $0.2$ &$1.0\times 10^{-10}$&{$1.0\times 10^{13}$}\\ 
\hline
\end{tabular}
\caption{List of experimental parameters used in numerical simulations. Here $p_d$ is the dark counting rate per pulse of Charlie's detectors; $e_d$ is the misalignment-error probability; $\eta_d$ is the detection efficiency of Charlie's detectors; $f$ is the error correction inefficiency; $\alpha_f$ is the fiber loss coefficient ($dB/km$); $\xi$ is the failure probability while using Chernoff bound~\cite{chernoff1952measure}; $N$ is the number of total pulse pairs sent out in the protocol.}\label{exproperty}
\end{table*}

\begin{figure}
\centering
\includegraphics[width=9cm]{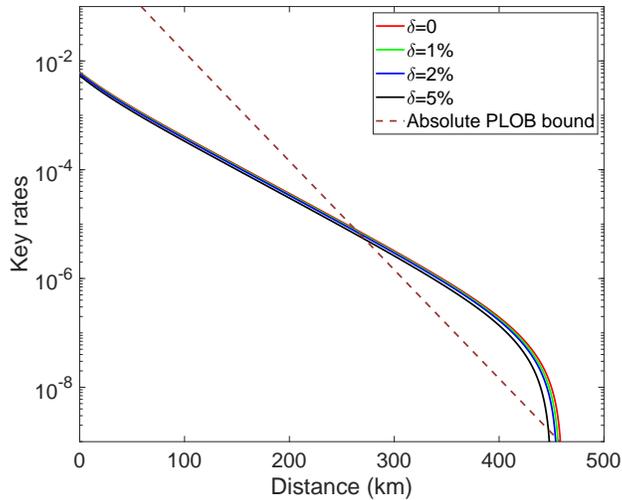}
\caption{The the key rates of the original SNS protocol under different degrees of intensity fluctuation. The absolute PLOB bound is the PLOB bound with $100\%$ detection efficiency detectors. Here we assume the symmetric channel, i.e., the distance between Alice and Charlie, $L_{AC}$, equals to the distance between Bob and Charlie, $L_{BC}$. The experiment parameters are listed in Table.~\ref{exproperty}.}\label{original}
\end{figure}

\begin{figure}
\centering
\includegraphics[width=9cm]{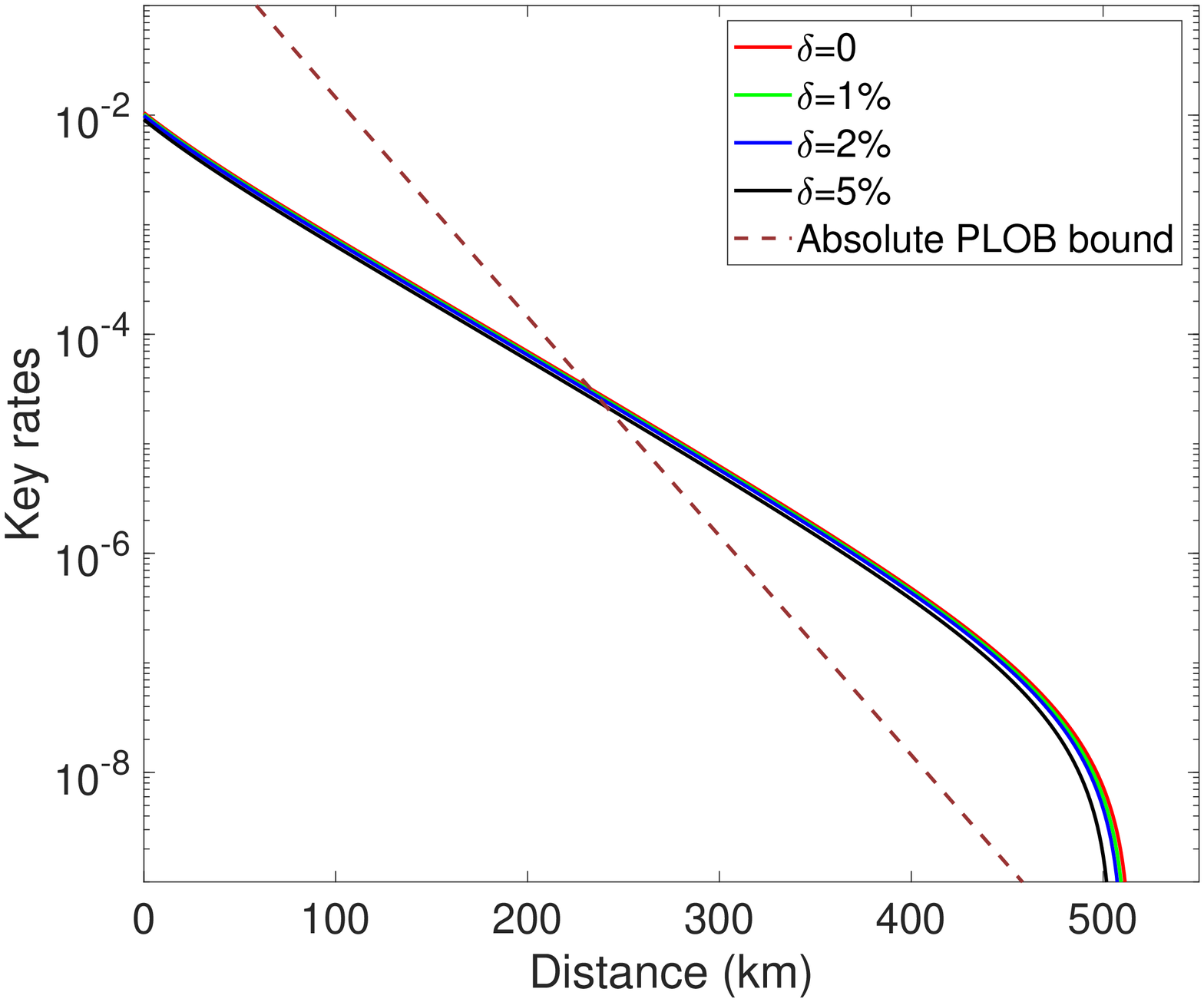}
\caption{The the key rates of the SNS protocol with AOPP under different degrees of intensity fluctuation. The absolute PLOB bound is the PLOB bound with $100\%$ detection efficiency detectors. Here we assume the symmetric channel, i.e., the distance between Alice and Charlie, $L_{AC}$, equals to the distance between Bob and Charlie, $L_{BC}$. The experiment parameters are listed in Table.~\ref{exproperty}.}\label{AOPP}
\end{figure}

\begin{figure}
\centering
\includegraphics[width=9cm]{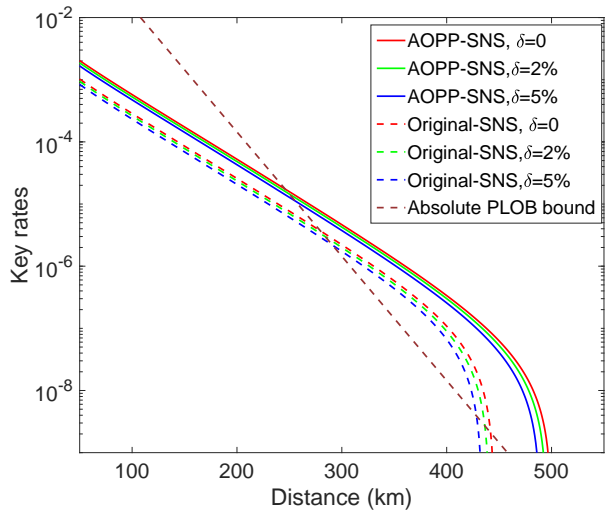}
\caption{The comparison of the key rates of the original SNS protocol and the SNS protocol with AOPP method under different degrees of intensity fluctuation. The absolute PLOB bound is the PLOB bound with $100\%$ detection efficiency detectors. Here we assume the asymmetric channel, where the distance between Alice and Charlie, $L_{AC}$, and the distance between Bob and Charlie, $L_{BC}$, satisfy $L_{AC}-L_{BC}=$50 km. The experiment parameters are listed in Table.~\ref{exproperty}.}\label{comp_curve}
\end{figure}

Figures~\ref{original} and \ref{AOPP} are the key rates of the original SNS protocol and the SNS protocol with the AOPP method under different degrees of intensity fluctuation. We assume a symmetric channel, which means that the distance between Alice and Charlie, $L_{AC}$, equals the distance between Bob and Charlie, $L_{BC}$. Figure~\ref {original} and \ref{AOPP} show that the key rates and farthest distance decrease slightly as the intensity fluctuation range increases. Figure~\ref{comp_curve} is the comparison of the key rates of the original SNS protocol and the SNS protocol with the AOPP method under different degrees of intensity fluctuation, where we set $L_{AC}-L_{BC} = $50 km. {The key rates at the distance of 350 km are shown in Table~\ref{comp_rate}, where the experiment parameters are the same as those of Figure~\ref{comp_curve}}. {In Figures~\ref{original}-\ref{comp_curve}, the absolute PLOB bound is for the case with perfect detection efficiency. Results in those Figures show that the key rates of our method can still exceed the absolute PLOB bound with $5\%$ intensity fluctuation. As the intensity fluctuation increases, so do the distance of the first intersection between the absolute PLOB bound and the key rate curve, implying that it is more difficult to exceed the absolute PLOB bound with a larger intensity fluctuation, and this trend is more apparent in Figure~\ref{comp_curve}. The key rates and distances corresponding to the first intersections in Figures~\ref{original}-\ref{comp_curve} are shown in Table~\ref{inter4}.} 

We use a more efficient method to calculate the key rates of the AOPP-SNS in Table~\ref{comp_rate}, by scanning the expected value of the counting rate of vacuum sources $\mean{S_{00}}$ in its range, which is

\begin{equation}
R^{\prime\prime}=\min_{\mean{S_{00}}}R^{\prime}(\mean{S_{00}}),
\end{equation}
and this can improve the non-asymptotic key rate a little bit. The results in Table~\ref{comp_rate} show that the key rates of AOPP with {$5\%$ intensity fluctuation are still higher than those of the} original SNS protocol without intensity fluctuation. With Table~\ref{comp_rate}, we can see that every $1\%$ intensity fluctuation causes a $4\%$ drop in key rate, which shows that the SNS protocol is robust against the source errors.

\begin{table*}[htbp]
\centering
\begin{tabular}{ccccccccc}
\hline
$\delta$ & $0\%$& $2\%$& $5\%$& $0\%$& $2\%$& $5\%$ \\ 
\hline
Channel & symmetric & symmetric& symmetric& symmetric& symmetric& symmetric\\
Method & AOPP & AOPP& AOPP & Original & Original& Original\\
Distance & $233$ & $237$& $241$ & $263$ & $267$& $272$\\
Key rate & $3.16\times 10^{-5}$ & $2.67\times 10^{-5}$& $2.18\times 10^{-5}$ &$8.00\times 10^{-6}$ & $6.66\times 10^{-6}$& $5.24\times 10^{-6}$\\
\hline
Channel & asymmetric & asymmetric& asymmetric& asymmetric& asymmetric& asymmetric\\
Method & AOPP & AOPP& AOPP & Original & Original& Original\\
Distance & $246$ & $250$& $256$& $280$ & $284$ & $291$\\
Key rate & $1.73\times 10^{-5}$ & $1.44\times 10^{-5}$& $1.11\times 10^{-5}$& $3.66\times 10^{-6}$ & $3.00\times 10^{-6}$& $2.18\times 10^{-6}$ \\
\hline
\end{tabular}
\caption{The key rates and distances corresponding to the first intersections between the absolute PLOB bound and the key rate curves in Figures~\ref{original}-\ref{comp_curve}. Here the results of the original method with symmetric channel are the intersections in Figure~\ref{original}; the results of the AOPP method with symmetric channel are the intersections in Figure~\ref{AOPP}; the results of the AOPP and original methods with asymmetric channel are the intersections in Figure~\ref{comp_curve}.}\label{inter4}
\end{table*}

\begin{table*}[htbp]
\centering
\begin{tabular}{ccccc}
\hline
$\delta$& $0\%$&$2\%$&$5\%$ & $10\%$\\
\hline
Original-SNS& $5.8\times 10^{-7}$& $5.15\times 10^{-7}$ & $4.35\times 10^{-7}$ & $3.26\times 10^{-7}$ \\ 
\hline
AOPP-SNS &$1.33\times 10^{-6}$ &$1.20\times 10^{-6}$& $1.05 \times 10^{-6}$ & $8.46\times 10^{-7}$\\
\hline
\end{tabular}
\caption{The key rates of the original SNS protocol and the AOPP method under different degrees of intensity fluctuation. The distance between Alice and Bob is 350 km, and the experiment parameters are the same with those of Figure.~\ref{comp_curve}. In our simulation, the fluctuation parameter $\delta$ is a range between $-\delta$ and $\delta$. For example, $\delta=10\%$ means a fluctuating range of $20\%$, from $-10\%$ to $+10\%$.}\label{comp_rate}
\end{table*}

The AOPP method was used in the SNS experiment in 511 km field fiber to achieve a higher key rate~\cite{chen2021twin}. We calculate the key rate of this experiment if there are source errors using the experiment observed values shown in Ref.~\cite{chen2021twin}, and the results show that the key rate is still positive even if $\delta$ is as large as $9.5\%$.

{In the security proof and numerical simulation above, we assume the intensities of the pulses are in a certain interval. But a more practical case is that very few pulses can exceed the bound values and we can only determine at most how many pulses outside the interval with a small failure probability. We show how to calculate the key rate under this case in Appendix~\ref{sns_s}.}

\section{Conclusion}
In this study, we propose a strict method for calculating the key rate of the SNS protocol with source errors, by establishing the equivalent protocols {through virtual attenuation and tagged model.} We finally obtain the key rate formulas under the premise of ensuring the protocol’s security. Our method can be combined with the AOPP method to further improve the key rate. The numerical results show that every $1\%$ intensity fluctuation causes a $4\%$ drop in key rate; additionally, the SNS protocol’s farthest distance slightly decreases as the intensity fluctuation range increases, indicating that the SNS protocol is robust against the source errors. {Our method can be directly applied to the 3-intensity SNS protocol with source error where `3-intensity' refers $\mu_{a_2} = \mu_{a_z},\mu_{b_2} = \mu_{b_z}$.}

{Since our method on the one hand allows Eve to know the error values of candidate pulses in advance, on the other hand requests no information leakage of state choice, the errors must be independent of state choice in applying our method~\cite{mizutani2019quantum}. With this condition being respected, our method applies to errors inside the bound values with whatever patterns. Our method does not work with crossing correlations between intensity errors and state choice at different time windows, say, setting-choice-dependent errors~\cite{pereira2020quantum,zapatero2021security,sixto2022security}, because such types of crossing correlations obviously lead to the state-choice information leakage to Eve if she knows the error values in advance. It is an interesting problem for future study on the setting-choice-dependent errors~\cite{pereira2020quantum,zapatero2021security,sixto2022security} with TF-QKD protocols.}

\textbf{Funding.} This work was supported by Ministration of Science and Technology of China through The National Key Research and Development Program of China Grant No. 2020YFA0309701; National Natural Science Foundation of China Grant Nos. 12174215, 12104184, 11774198 and 11974204; Shandong Provincial Natural Science Foundation Grant No. ZR2021LLZ007; Key R$\&$D Plan of Shandong Province Grant No. 2021ZDPT01; Open Research Fund Program of the State Key Laboratory of Low-Dimensional Quantum Physics Grant No. KF202110; Leading Talents of Quancheng Industry.

\appendix
\section{The calculation method}\label{cal}
Except the data of the $Z$ windows, the other data are used to perform the decoy state analysis including the data when either or both of Alice and Bob choose the decoy windows. In the decoy-state analysis, we shall only focus on these data (all data except the data of the $Z$ windows). 

We simplify the symbols of  two pulse sources $a_l,b_r(l,r=0,1,2)$ as $lr$. We denote the number of pulse pairs of source $lr$ sent out in the whole protocol by $N_{lr}$, and the total number of one-detector heralded events of source $lr$ by $n_{lr}$, {whose expected value is $\mean{n_{lr}}$}. We define the counting rate of source $lr$ by $S_{lr}=n_{lr}/N_{lr}$, and the corresponding expected value by $\mean{S_{lr}}$. The Chernoff bound can be used to estimate the lower and upper bound of the expected values according to their observed values. We have
\begin{equation}
\begin{split}
N_{00}=&[(1-p_{a_z})(1-p_{b_z})p_{a_0}p_{b_0}+(1-p_{a_z})p_{b_z}p_{a_0}(1-\epsilon_b)\\
&+p_{a_z}(1-p_{b_z})(1-\epsilon_a)p_{b_0}]N,\\
N_{01}=&[(1-p_{a_z})p_{a_0}+p_{a_z}(1-\epsilon_a)](1-p_{b_z})p_{b_1}N,\\
N_{10}=&[(1-p_{b_z})p_{b_0}+p_{b_z}(1-\epsilon_b)](1-p_{a_z})p_{a_1}N,\\
N_{02}=&[(1-p_{a_z})p_{a_0}+p_{a_z}(1-\epsilon_a)](1-p_{b_z})p_{b_2}N,\\
N_{20}=&[(1-p_{b_z})p_{b_0}+p_{b_z}(1-\epsilon_b)](1-p_{a_z})p_{a_2}N.
\end{split}
\end{equation}

We denote the number of pulse pairs of the $X$ windows sent out in the whole protocol by $N_{X}$, and the number of effective wrong events by $M_{X}$, whose expected value is $\mean{M_{X}}$. We define $T_X=M_X/N_X$, and its expected value by $\mean{T_X}$.

Since the sources are unstable, we can not directly apply the results of traditional decoy-state method~\cite{wang2005beating,yu2019sending,hu2019general}. As the  discussion in the main body, we can take the virtual protocol 2 to calculate the final key rate according to the observed values in the real protocol. The density matrices of the stable sources $a_1^\prime$ and $b_1^\prime$ are
\begin{equation}
\begin{split}
&\rho_{a_1^\prime}=\sum_{k=0}^{+\infty}c_{a_1}^k\oprod{k}{k},\quad c_{a_1}^k=\frac{(\mu_{a_1}^U)^ke^{-\mu_{a_1}^U}}{k!},\\
&\rho_{b_1^\prime}=\sum_{k=0}^{+\infty}c_{b_1}^k\oprod{k}{k},\quad c_{b_1}^k=\frac{(\mu_{b_1}^U)^ke^{-\mu_{b_1}^U}}{k!}.
\end{split}
\end{equation}
For the unstable sources $a_2^\prime$ and $b_2^\prime$, the density matrices of the pulses in the $i-$th time window are
\begin{equation}
\begin{split}
&\rho_{a_2^\prime}^i=\sum_{k=0}^{+\infty}c_{a_2}^{k,i}\oprod{k}{k},\quad c_{a_2}^{k,i}=\frac{(\mu_{a_2}^{\prime i})^ke^{-\mu_{a_2}^{\prime i}}}{k!},\\
&\rho_{b_2^\prime}^i=\sum_{k=0}^{+\infty}c_{b_2}^{k,i}\oprod{k}{k},\quad c_{b_2}^{k,i}=\frac{(\mu_{b_2}^{\prime i})^ke^{-\mu_{b_2}^{\prime i}}}{k!}.
\end{split}
\end{equation}

For the stable two pulse sources, we denote the expected values of the counting rate of the state $\oprod{01}{01}$ and $\oprod{10}{10}$ as $\mean{s_{01}}$ and $\mean{s_{10}}$. We have~\cite{wang2008general}
\begin{equation}
\mean{s_{01}}\ge\mean{s_{01}}^L=\frac{c_{b_2}^{2,L}\mean{S_{01}}^L-c_{b_1}^2\mean{S_{02}}^U-(c_{b_1}^0c_{b_2}^{2,L}-c_{b_1}^2c_{b_2}^{0,L})\mean{S_{00}}^U}{c_{b_1}^1c_{b_2}^{2,L}-c_{b_1}^2c_{b_2}^{1,L}},
\end{equation}  
and
\begin{equation}
\mean{s_{10}}\ge\mean{s_{10}}^L=\frac{c_{a_2}^{2,L}\mean{S_{10}}^L-c_{a_1}^2\mean{S_{20}}^U-(c_{a_1}^0c_{a_2}^{2,L}-c_{a_1}^2c_{a_2}^{0,L})\mean{S_{00}}^U}{c_{a_1}^1c_{a_2}^{2,L}-c_{a_1}^2c_{a_2}^{1,L}}.
\end{equation} 
{If Bob does the error correction computation and he chooses to take decoy-state analysis after error correction, he can verify the bounds of $\mean{S_{00}}$ more tightly by using the events of all those time windows $a_0b_0$, i.e., when both parties used vacuum source, because he knows all of them after error correction. Say, after error correction, for all those time windows of survived pairs in AOPP, he knows which bits are from $a_0b_0$, while for other time windows including those time windows of the rejected pairs and those time windows not participating in the pairing process of AOPP, they can publicly announce the corresponding intensities of pulses they each have used. In this way Bob can simply use $n_{a_0b_0}/N_{a_0b_0}$ for his observed value of $\mean{S_{00}}$, and hence the protocol does not need to reserve any vacuum windows as random samples to test $S_{00}$. This improves the non-asymptotic key rate a little bit because we can verify $\mean{S_{00}}$ more efficiently.}

We denote the expected values of the counting rate of the untagged pulses as $\mean{s_{1}}$, and we have~\cite{hu2019general}
\begin{equation}
\mean{s_{1}}\ge \mean{s_{1}}^L=\frac{\mu_{a_1}^U}{\mu_{a_1}^U+\mu_{b_1}^U}\mean{s_{10}}^L+\frac{\mu_{b_1}^U}{\mu_{a_1}^U+\mu_{b_1}^U}\mean{s_{01}}^L.
\end{equation} 
The upper bound of the expected value of the phase-flip error rate is
\begin{equation}
\mean{e_1^{ph}}\le \mean{e_1^{ph}}^U=\frac{\mean{T_X}^U-e^{-\mu_{a_1}^U-\mu_{b_1}^U}\mean{S_{00}}^L/2}{e^{-\mu_{a_1}^U-\mu_{b_1}^U}(\mu_{a_1}^U+\mu_{b_1}^U)\mean{s_{1}}^L}. 
\end{equation}

Then we have the lower bound of the real value of the untagged bits in the $Z$ windows
\begin{equation}
n_1^L=O^L(\mean{n_1}^L)
\end{equation}
where
\begin{equation}
\mean{n_1}^L=Np_{a_z}p_{b_z}[\epsilon_a(1-\epsilon_b)c_{a_z}^1+\epsilon_b(1-\epsilon_a)c_{b_z}^1]\mean{s_{1}}^L,
\end{equation}
and $O^L(Y)$ is defined in Eq.~\eqref{OL}. And the upper bound of the real value of the phase-flip error rate is
\begin{equation}
e_1^{ph,U}=\frac{O^U(n_1^L\mean{e_1^{ph}}^U)}{n_1^L}.
\end{equation}

Finally, we get the secure final key rate
\begin{equation}
\begin{split}
R=\frac{1}{N}\left\{n_1^L[1-H(e_1^{ph,U})]-fn_tH(E)-\log_2 \frac{2}{\varepsilon_{cor}}-2\log_2\frac{1}{\sqrt{2}\varepsilon_{PA}\hat{\varepsilon}}\right\}.
\end{split}
\end{equation}
We set $\varepsilon_{cor}=\varepsilon_{PA}=\hat{\varepsilon}=10^{-10}$ in this article. 

Besides, with all those data, we can use the AOPP method shown in Ref.~\cite{jiang2021composable} to get higher key rate. With values above, we can calculate the lower bound of untagged bits and phase-flip error rate after AOPP, $n_1^{\prime L}$ and $e_{1}^{\prime ph,U}$ by the method proposed in Refs.~\cite{jiang2020zigzag,jiang2021composable}. We have the related formulas of $n_1^{\prime,L}$ as follows:
\begin{subequations}
\begin{align}
&\mean{{n_{u1}}}^L=Np_{a_z}p_{b_z}\epsilon_a(1-\epsilon_b)c_{a_z}^1\mean{s_{10}}^L,\\
&\mean{{n_{u0}}}^L=Np_{a_z}p_{b_z}\epsilon_b(1-\epsilon_a)c_{b_z}^1\mean{s_{01}}^L,\\
&u=\frac{n_g}{2n_{odd}},\\
&n_{u1}^L=\varphi^L(u\mean{{n_{u1}}}^L),\\
&n_{u0}^L=\varphi^L(u\mean{{n_{u0}}}^L),\\
&n_{1}^L=n_{u1}^L+n_{u0}^L,\\
&n_1^r=\varphi^L\left(\frac{(n_{1}^L)^2}{2un_t}\right),\\
&n_{u1}^\prime=2n_1^r\left(\frac{n_{u1}^L}{n_{1}^L}-\sqrt{-\frac{\ln\varepsilon}{2n_1^r}}\right)\\
&n_{u0}^\prime=2n_1^r\left(\frac{n_{u0}^L}{n_{1}^L}-\sqrt{-\frac{\ln\varepsilon}{2n_1^r}}\right)\\
&n_{min}=\min(n_{01}^\prime,n_{10}^\prime),\\
&n_1^{\prime,L}=2\varphi^L\left(n_{min}(1-\frac{n_{min}}{2n_1^r})\right),
\end{align}
\end{subequations}
where $n_g$ is the number of pair if Alice and Bob perform AOPP to their raw keys; ${n_{odd}}$ is the number of pairs with odd-parity if Bob randomly groups all the raw keys two by two, and $n_g$ and $n_{odd}$ are observed values; $\varepsilon$ is the failure probability of parameter estimation; and $\varphi^U(x),\varphi^L(x)$ are the upper and lower bounds while using Chernoff bound~\cite{chernoff1952measure} to estimate the real values according to the expected values, whose details are shown in Appendix.~\ref{chernoff}.

And we have the the related formulas of $e_{1}^{\prime ph}$ as follows:
\begin{subequations}
\begin{align}
&r=\frac{n_1^L}{n_1^L-2n_1^r}\ln\frac{3(n_1^L-2n_1^r)^2}{\varepsilon},\\
&e_{\tau}=\frac{\varphi^U(2n_1^r\mean{{e_1^{ph}}}^U)}{2n_1^r-r}\\
&{M}_s^U=\varphi^U[(n_1^r-r){e_{\tau}}(1-{e_{\tau}})]+r,\\
&e_{1}^{\prime ph,U}=\frac{2{M}_s^U}{n_1^\prime}
\end{align}
\end{subequations} 

Finally, we have the following final key length formula after AOPP
\begin{equation}
\begin{split}
R^\prime=\frac{1}{N}\left\{n_1^{\prime L}[1-H(e_{1}^{\prime ph,U})]-fn_t^\prime H(E^\prime)-2\log_2{\frac{2}{\varepsilon_{cor}}}-4\log_2{\frac{1}{\sqrt{2}\varepsilon_{PA}\hat{\varepsilon}}}\right\}.
\end{split}
\end{equation}

\section{Chernoff bound}\label{chernoff}
The Chernoff bound can help us estimate the expected value from their observed values~\cite{chernoff1952measure}. Let $X_1,X_2,\dots,X_n$ be $n$ independent random samples, detected with the value 1 or 0, and let $X$ denote their sum satisfying $X=\sum_{i=1}^nX_i$. $E$ is the expected value of $X$. We have
\begin{align}
\label{EL}E^L(X)=&\frac{X}{1+\delta_1(X)},\\
\label{EU}E^U(X)=&\frac{X}{1-\delta_2(X)},
\end{align}
where we can obtain the values of $\delta_1(X)$ and $\delta_2(X)$ by solving the following equations
\begin{align}
\label{delta1}\left(\frac{e^{\delta_1}}{(1+\delta_1)^{1+\delta_1}}\right)^{\frac{X}{1+\delta_1}}&=\xi,\\
\label{delta2}\left(\frac{e^{-\delta_2}}{(1-\delta_2)^{1-\delta_2}}\right)^{\frac{X}{1-\delta_2}}&=\xi,
\end{align}
where $\xi$ is the failure probability.

Besides, we can use the Chernoff bound to help us estimate their real values from their expected values. Similar to Eqs.~\eqref{EL}- \eqref{delta2}, the observed value, $O$, and its expected value, $Y$, satisfy 
\begin{align}
\label{OU}&O^U(Y)=[1+\delta_1^\prime(Y)]Y,\\
\label{OL}&O^L(Y)=[1-\delta_2^\prime(Y)]Y,
\end{align}   
where we can obtain the values of $\delta_1^\prime(Y,\xi)$ and $\delta_2^\prime(Y,\xi)$ by solving the following equations
\begin{align}
\left(\frac{e^{\delta_1^\prime}}{(1+\delta_1^\prime)^{1+\delta_1^\prime}}\right)^{Y}&=\xi,\\
\label{endd}\left(\frac{e^{-\delta_2^\prime}}{(1-\delta_2^\prime)^{1-\delta_2^\prime}}\right)^{Y}&=\xi.
\end{align}

\section{Some remarks of our protocol and applicability of Chernoff bound}\label{decoy}
{In applying our method, we don't request the values of intensities (or intensity errors) to be random, there can be patterns among those data. There are two important points here: First, Alice and Bob does not need to know the values of intensities; second, we assume that Eve can know the values in advance.}
 
{Our method can be efficiently understood in this way: the intensity values of candidate pulses can be predetermined by any third party including Eve in whatever way provided that in predetermining them, he or she has no information of the secret state choice taken by Alice and Bob in the experiment, e.g., choosing which state to send out at each time windows~\cite{mizutani2019quantum}. In predetermining those explicit values of intensities (or intensity errors), the third party does not have to generate them randomly. He (or she) can take whatever way, e.g., he (or she) can first generate the values of a few time windows and then based on these values to determine all values by using certain functional. The condition of ``no information of the secret state choice taken by Alice and Bob'' for the third party only forbids the crossing correlation among intensity values and the secret state choice taken by Alice and Bob~\cite{mizutani2019quantum}, but it does not forbid other types of dependence of intensity values among different time windows.}

{We use the fact that there exist explicit values of intensity errors in a real experiment. If one regard this as a ``condition'' rather than a ``fact'', then this condition actually always hold in any real experiment. If Eve knows those intensity values explicitly, our virtual protocols with attenuations hold exactly and hence the result of our real protocol is secure, as shown in the text. If Eve does not know those values explicitly, the result of our real protocol must be also secure because a weakened Eve who does not know those values explicitly cannot be more powerful in attacking the real protocol than an unweakened Eve who knows them explicitly. In particular, our result holds even though the third party  predetermines the intensity errors in a certain probabilistic distribution of many sets of values of intensity errors of different time windows rather than one set of deterministic values. Because such a case simply means that the intensity values could be set 1, could be set 2, and so on. No matter which set it actually is, Alice and Bob can go ahead to obtain the their final bits by the real protocol because they do not need the explicit values in carrying out the real protocol. On the other hand, the final bits obtained by the real protocol must be also secure. Again, this is  because a weakened Eve who does not know explicitly which set of intensity values in the experiment cannot be more powerful than the Eve who knows it explicitly. }

{In this paper, we apply the decoy-state analysis proposed in Ref.~\cite{wang2008general} to estimate the parameters of SNS protocol. To show that the Chernoff bound can be applied here, we take the decoy-state polarization BB84 protocol as an example, where Alice take decoy-state method with 3 sources which are the vacuum source $o$, the decoy source $x$ and the signal source $y$. Given discussions above, we can now do the calculation with the following model for source intensity errors:}

{We assume that all values of intensities (or intensity errors) are in a known range~\cite{wang2008general}. There are no crossing independence between intensity values and the secret information of state choice~\cite{mizutani2019quantum}.} With these conditions, we take the worst-case that Eve knows in advance the intensity errors explicitly of each candidate pulses. In our study, we use constant probabilities $p_o, p_x, p_y$ for Alice to choose the vacuum source $o$, the decoy source $x$, and the signal source $y$ at each time windows. Equivalently, we can use the following product state~\cite{mizutani2019quantum} for the source in our model:
\begin{equation}
\rho_{AB}=\otimes_{i=1}^N(p_o\rho_o\otimes\oprod{o_i}{o_i}_A+p_x\rho_{\mu_{x_i}}\otimes\oprod{x_i}{x_i}_A+p_y\rho_{\mu_{y_i}}\otimes\oprod{y_i}{y_i}_A)
\end{equation}
where the subscript $A$ represents Alice's classical memories of storing which source are chosen, the states $\rho_o,\rho_{\mu_{x_i}},\rho_{\mu_{y_i}}$ are send out to the channel. We have
\begin{align}
\label{rho0}&\rho_o=\sum_{k=0}a_{o_i}^k\oprod{k}{k},\quad a_{o_i}^k=\delta_{0,k},\\
&\rho_{\mu_{x_i}}\sum_{k=0}a_{{x_i}}^k\oprod{k}{k},\quad a_{{x_i}}^k=\frac{(\mu_{x}^i)^ke^{-{\mu_{x}^i}}}{k!},\\
\label{rhoy}&\rho_{\mu_{y_i}}\sum_{k=0}a_{{y_i}}^k\oprod{k}{k},\quad a_{{y_i}}^k=\frac{(\mu_{y}^i)^ke^{-{\mu_{y}^i}}}{k!}.
\end{align}
We assume $\mu_{x}^L\le \mu_{x}^i\le \mu_{x}^U, \mu_{y}^L\le \mu_{y}^i\le \mu_{y}^U$ and $\mu_{x}^L,\mu_{x}^U,\mu_{y}^L,\mu_{y}^U$ are known values. This assumption is the same as that of Ref.~\cite{wang2008general}.

With Eqs.(\ref{rho0}-\ref{rhoy}), we can rewrite $\rho_{AB}$ as the following form
\begin{equation}\label{rabd}
\rho_{AB}=\otimes_{i=1}^N\left[\sum_{k=0}\left(p_oa_{o_i}^k\oprod{o_i}{o_i}_A+p_xa_{{x_i}}^k\oprod{x_i}{x_i}_A+p_ya_{{y_i}}^k\oprod{y_i}{y_i}_A\right)\otimes
\oprod{k}{k}\right].
\end{equation} 
Eq.~\eqref{rabd} shows that the actual state for the sent-out pulse of time window $i$ is simply an eigenstate in Fock space, with a classical probability for different photon numbers. Thus we can always regards that the actual state for the pulse to be sent out in the $i$-th time window is in a certain photon-number state $\oprod{m_i}{m_i}$. Similar idea is also used in the original decoy-state method with exact intensities. At any time, the source emits a phase randomized coherent state which can be written in the probabilistic mixture of different photon-number state. We imagine that it actually emits a Fock state at each time window, with the specific photon number being chosen by the corresponding probability distribution. Thus we can take the state prepared by Alice as
\begin{equation}\label{eq28}
\tilde{\rho}_{AB}=\otimes_{i=1}^N\left[\left(p_oa_{o_i}^{m_i}\oprod{o_i}{o_i}_A+p_xa_{{x_i}}^{m_i}\oprod{x_i}{x_i}_A+p_ya_{{y_i}}^{m_i}\oprod{y_i}{y_i}_A\right)\otimes
\oprod{m_i}{m_i}\right].
\end{equation}   
This is a product state of Alice's local state and set-out state as following:
\begin{equation}
\tilde{\rho}_{AB}=\rho_{\mathcal{L}}\otimes\rho_{\mathcal{T}}.
\end{equation}
Here either the local state $\rho_{\mathcal{L}}$ or the sent-out state $\rho_{\mathcal{T}}$ itself is a product state of different time windows:
\begin{align}
\label{local}&\rho_{\mathcal{L}}=\otimes_{i=1}^N\left(p_oa_{o_i}^{m_i}\oprod{o_i}{o_i}_A+p_xa_{{x_i}}^{m_i}\oprod{x_i}{x_i}_A+p_ya_{{y_i}}^{m_i}\oprod{y_i}{y_i}_A\right),\\
&\rho_{\mathcal{T}}=\otimes_{i=1}^N  \oprod{m_i}{m_i}
\end{align}
Surely, any Eve's attack, including the attacks to the sent-out pulses can never change the local state $\rho_{\mathcal{L}}$, if Eve has no access to Alice's lab. If the pulse in $i$-th window causes a click, Alice can measure her local state to determine which source the pulse belongs to, and the probability that it is from source $l,l=o,x,y$ is
\begin{equation}
p_{i,l}^{m_i}=\frac{p_la_{l_i}^{m_i}}{p_oa_{o_i}^{m_i}+p_xa_{{x_i}}^{m_i}+p_ya_{{y_i}}^{m_i}}.
\end{equation}
Given the product form of local state shown in Eq.~\eqref{local}, observing $\rho_{\mathcal{L}_i}=p_oa_{o_i}^{m_i}\oprod{o_i}{o_i}_A+p_xa_{{x_i}}^{m_i}\oprod{x_i}{x_i}_A+p_ya_{{y_i}}^{m_i}\oprod{y_i}{y_i}_A$ does not affect any state $\rho_{\mathcal{L}_j}=p_oa_{o_j}^{m_j}\oprod{o_j}{o_j}_A+p_xa_{{x_j}}^{m_j}\oprod{x_j}{x_j}_A+p_ya_{{y_j}}^{m_j}\oprod{y_j}{y_j}_A$, provided that $i\neq  j$. Thus we have {\em Fact C1}: values of $p_{i,l}^{m_i}$ of different time windows are {\em independent}.

Similar to Eq. (33) in Ref.~\cite{wang2008general}, we have the following asymptotic formulas for the number of counts caused by each kind of pulses
\begin{equation}
\mean{n_k^l}=\sum_{i\in c_k} p_{i,l}^{k}=\sum_{i\in c_k}\frac{p_la_{l_i}^{k}}{p_oa_{{o_i}}^{k}+p_xa_{{x_i}}^{k}+p_ya_{{y_i}}^{k}},
\end{equation}
where $\mean{n_k^l}$ is the expected value of the counts cased by the k-photon state from source $l$, and $c_k$ is the set of all windows that cause counts by k-photon state pulses. We also have the expected value of all counts cased by source $l$
\begin{equation}\label{nl}
\mean{N_l}=\sum_{k=0} \mean{n_k^l}=\sum_{k=0} \sum_{i\in c_k} p_{i,l}^{k}. 
\end{equation} 
Since all $p_{i,l}^{k}$ are independent, we can apply the Chernoff bound to estimate the bound values of $\mean{N_l}$ according to its corresponding observed values. 

With Eq.~\eqref{nl} for $l=o,x,y$, we can get the lower bound of $\mean{n_1^y}$~\cite{wang2008general}
\begin{equation}\label{n1y}
\mean{n_1^y}\ge \mean{n_1^y}^L=\frac{a_y^{1,L}\left[a_y^{2,L}\mean{N_x}^L-a_x^{2,U}\mean{N_y}^U-(a_y^{2,L}a_x^{0,U}-a_x^{2,U}a_y^{0,L})\mean{N_0}^Up_y/p_o\right]}{a_x^{1,U}a_y^{2,L}-a_y^{1,L}a_x^{2,U}},
\end{equation}
where the subscript $U,L$ represent the upper and lower bounds respectively.

Recall the express form of $\mean{n_1^y}$,
\begin{equation}
\mean{n_1^y}=\sum_{i\in c_1} p_{i,y}^{1}.
\end{equation} 
According to {\em Fact C1} above, all values $p_{i,y}^{1}$ of different $i$ in the summation above are independent, and hence we can use the Chernoff bound to estimate the lower bound of $n_1^y$ according to $\mean{n_1^y}$, where $n_1^y$ is the real value of the counts cased by the single-photon state from source $y$. We have
\begin{equation}
n_1^y\ge O^L(\mean{n_1^y}),
\end{equation} 
except for a failure probability $\xi$, where $O^L(Y)$ is defined in Eq.~\eqref{OL}. $O^L(Y)$ is a an increasing function of $Y$, which means if $Y_1\le Y_2$, then $O^L(Y_1)\le O^L(Y_2)$. Thus we have 
\begin{equation}
n_1^y\ge n_1^{y,L},
\end{equation}
except for a failure probability $4\xi$, where
\begin{equation}
n_1^{y,L}=O^L(\mean{n_1^y}^L).
\end{equation}

Above we have shown the worst-case result for the deterministic intensities of source pulses. Mathematically, this means
\begin{equation}
\Pr(n_1^y\ge n_1^{y,L}|\rho_D)\le 4\xi,
\end{equation}
where $\rho_D$ is classical memories that store the intensities of all pulse. Since we only use the intensity error range in the whole precess, the result obviously holds for the case of probabilistic intensities given the same intensity error range, which is saying
\begin{equation}
\Pr(n_1^y\ge n_1^{y,L})=\sum_{\rho_D}\Pr(n_1^y\ge n_1^{y,L}|\rho_D)\Pr(\rho_D) \le 4\xi.
\end{equation}

Above we have demonstrated the applicability of Chernoff bound through the decoy-state BB84 as an example. The same conclusion on the applicability obviously holds for SNS protocol which has been calculated in our main text.

\section{Calculation formulas with SNS protocol}\label{sns_s}
In the main text, we assume the intensities of the pulses are in a certain interval, $[(1-\delta)\mu,[(1+\delta)\mu]$. For a more practical case that we can only determine this interval probabilistically with a certain degree of confidence~\cite{mizutani2019quantum}, we can use the following method to calculate the key rate.

In the real protocol, denote $\mathcal{C}^I$ as a set containing all time windows in which the intensities of all candidate states are in a certain interval defined in Eq.(1) in the main text. Denote $\mathcal{C}^O$ as a set containing all time windows in which at least on candidate state's intensity is out of the certain intervals. Apparently, we can apply the virtual attenuation method proposed in the main text to the pulse pairs in $\mathcal{C}^I$ and we get the virtual protocol. As for the pulse pairs in $\mathcal{C}^O$, they are the same in the virtual protocol and the real protocol.

In the virtual protocol, we have
\begin{equation}
\begin{split}
&\rho_{a_l^\prime}^i=\sum_{k=0}^{+\infty}c_{a_l}^{k,i}\oprod{k}{k},\quad \rho_{b_r^\prime}^i=\sum_{k=0}^{+\infty}c_{b_r}^{k,i}\oprod{k}{k},
\end{split}
\end{equation}  
for $l,r=0,1,2,z$.

Let set $c_k$ contains all $\ket{k0}$-photon-pair one-detector heralded events caused by sources $(a_0^\prime b_0^\prime,a_1^\prime b_0^\prime,a_2^\prime b_0^\prime,a_z^\prime b_0^\prime)$. Let set $c_k^I$ be a subset of $c_k$ that contains all one-detector heralded events caused by pulse pairs in time windows of $\mathcal{C}^I$, and let set $c_k^O$ be a subset of $c_k$ that contains all one-detector heralded event caused by pulse pairs in time windows of $\mathcal{C}^O$. We have
\begin{align}
\label{n00}\mean{n_{00}}&=\sum_{i\in c_0}[p_{a_0}p_{b_0}+p_{a_0}p_{b_z}(1-\epsilon_b)+p_{a_z}(1-\epsilon_a)p_{b_0}]D_0^i \nonumber \\
&=\sum_{i\in c_0^I}[p_{a_0}p_{b_0}+p_{a_0}p_{b_z}(1-\epsilon_b)+p_{a_z}(1-\epsilon_a)p_{b_0}]D_0^i+\Delta_0,\\
\mean{n_{10}}&=\sum_{k=0}\sum_{i\in c_k}p_{a_1}[p_{b_0}+p_{b_z}(1-\epsilon_b)]c_{a_1}^{k,i}D_k^i \nonumber \\
&=\sum_{k=0}\sum_{i\in c_k^I}p_{a_1}[p_{b_0}+p_{b_z}(1-\epsilon_b)]c_{a_1}^{k,i}D_k^i +\Delta_1,\\
\mean{n_{20}}&=\sum_{k=0}\sum_{i\in c_k}p_{a_2}[p_{b_0}+p_{b_z}(1-\epsilon_b)]c_{a_2}^{k,i}D_k^i \nonumber \\
&=\sum_{k=0}\sum_{i\in c_k^I}p_{a_2}[p_{b_0}+p_{b_z}(1-\epsilon_b)]c_{a_2}^{k,i}D_k^i+\Delta_2,
\end{align}
where
\begin{equation}
D_k^i=\frac{1}{[p_{b_0}+p_{b_z}(1-\epsilon_b)]\{[p_{a_0}+p_{a_z}(1-\epsilon_a)]c_{a_0}^{k,i}+p_{a_1}c_{a_1}^{k,i}+p_{a_2}c_{a_2}^{k,i}+p_{a_z}\epsilon_a c_{a_z}^{k,i}\}}.
\end{equation}

In this case, we determine the untagged bits only from the sources $a_0b_z^{\prime 1}$ and $a_z^{\prime 1}b_0$ of $Z$ windows in $\mathcal{C}^I$, so we have
\begin{equation}\label{nu1sum}
\mean{n_{u1}}=\sum_{i\in c_1^I} p_{a_z}\epsilon_ap_{b_z}(1-\epsilon_b)c_{a_z}^1 D_1^i.
\end{equation}

Denote there are $n_{\Delta}$ elements in set $\mathcal{C}^O$. Similar to Ref.~\cite{mizutani2019quantum}, we assume $n_{\Delta}$ is upper bounded by $N_{\Delta}$ with a failure probability at most $\varepsilon_{fail}$. Using the fact that there are at most $N_{\Delta}$ one-detector heralded event caused by pulse pairs in time windows of $\mathcal{C}^O$, we have
\begin{equation}\label{eqdelta}
0\le \Delta_{j}\le N_{\Delta}, \quad j=0,1,2.
\end{equation}

Combining Eqs.(\ref{n00}-\ref{eqdelta}), we have
\begin{equation}
\mean{{n_{u1}}}^L=Np_{a_z}p_{b_z}\epsilon_a(1-\epsilon_b)c_{a_z}^1\mean{s_{10}}^{\prime,L},
\end{equation}
where
\begin{equation}\label{s100}
\mean{s_{10}}^{\prime,L}=\frac{c_{a_2}^{2,L}[\mean{S_{10}}^L-N_{\Delta}/N_{10}]-c_{a_1}^2\mean{S_{20}}^U-(c_{a_1}^0c_{a_2}^{2,L}-c_{a_1}^2c_{a_2}^{0,L})\mean{S_{00}}^U}{c_{a_1}^1c_{a_2}^{2,L}-c_{a_1}^2c_{a_2}^{1,L}}.
\end{equation}

Similarly, we can get the formulas of $\mean{{n_{u0}}}^L$ and $\mean{s_{01}}^{\prime,L}$.

Define 
\begin{equation}
\mean{s_{1}}^{\prime,L}=\frac{\mu_{a_1}^U}{\mu_{a_1}^U+\mu_{b_1}^U}\mean{s_{10}}^{\prime,L}+\frac{\mu_{b_1}^U}{\mu_{a_1}^U+\mu_{b_1}^U}\mean{s_{01}}^{\prime,L},
\end{equation}
we have the lower bound of the expected value of untagged bits
\begin{equation}
\mean{n_1}^L=Np_{a_z}p_{b_z}[\epsilon_b(1-\epsilon_a)c_{b_z}^1+\epsilon_a(1-\epsilon_b)c_{b_z}^1]\mean{s_{1}}^{\prime,L}.
\end{equation}

Denote the number of wrong untagged bits in the $X$ windows of $\mathcal{C}^I$ by $m_1$, whose corresponding expected value is $\mean{m_1}$, we have ${m_1}\le {M_X}$. The upper bound of the expected value of phase flip error rate of the untagged bits in the $Z$ windows of $\mathcal{C}^I$ satisfy
\begin{equation}\label{e1ph1}
\mean{e_1^{ph}}^U=\frac{\mean{T_X}^U}{e^{-\mu_{a_1}^U-\mu_{b_1}^U}(\mu_{a_1}^U+\mu_{b_1}^U)\mean{s_{1}}^{\prime,L}}. 
\end{equation}
Here we use the fact that the expected values of the untagged bits in the $X$ windows and $Z$ windows are proportional to their corresponding sending probabilities in the virtual protocol.

With all those values, we can calculate the final key rate of the SNS protocol. The failure probability $\varepsilon_{fail}$ to estimate the upper bound of $N_\Delta$ contributes to the failure probability of parameter estimation. The only differences of the formulas here and those in Appendix~\ref{cal} are the extra $N_{\Delta}/N_{10}$ in Eq.~\eqref{s100} and taking the lower bounds of vacuum counts as $0$ in Eq.~\eqref{e1ph1}. In the typical experiments of TF-QKD, the counting rate of vacuum pulses is about $10^{-8}$ or smaller, and taking it as $0$ has little effect on the key rate. Thus when $N_\Delta$ is small, for example $N_\Delta\sim 100$, we expect the method in this section would just affect the key rates a little.  

Apparently, we can introduce virtual local state similar to Eq.~\eqref{eq28} in deriving Eqs.~(\ref{n00}-\ref{nu1sum}). This means that quantities inside the summation of Eqs.~(\ref{n00}-\ref{nu1sum}) are independent to each other and hence the Chernoff bound applies in the calculation with finite data size.

\bibliography{refs.bib}

\end{document}